\documentclass[9pt,twocolumn,twoside]{opticajnl}
\journal{opticajournal} 

\setboolean{shortarticle}{false}

\usepackage{lineno}
\usepackage[capitalise]{cleveref}
\usepackage{algorithm}
\usepackage{algpseudocode}
\usepackage{booktabs}
\usepackage{makecell}
\usepackage{array}
\usepackage{multirow}
\usepackage{siunitx} 
\usepackage{threeparttable}

\title{Real-Time Megapixel Kilohertz Neuromorphic Shack-Hartmann Wavefront Sensor}

\author[1]{Yuhan Bao}
\author[1]{Chenxin Shao}
\author[1,2,*]{Kaiwei Wang}

\affil[1]{National Research Center for Optical Instrumentation, Zhejiang University, Hangzhou 310027, China}
\affil[2]{Central Research Institute of Sunny Optical Technology, Hangzhou 311200, China}

\affil[*]{wangkaiwei@zju.edu.cn}
\affil[]{ORCID: Yuhan Bao, 0009-0000-3833-4565; Kaiwei Wang, 0000-0002-8272-3119}

\begin{abstract}
Conventional frame-based Shack--Hartmann wavefront sensors (SHWFS) are limited by dynamic range and the intrinsic trade-off between spatial and temporal resolution, while high-bandwidth acquisition poses additional challenges for real-time wavefront reconstruction. 
This work presents a real-time, megapixel, kilohertz neuromorphic SHWFS to overcome these limitations. 
In static optical metrology, the proposed pipeline achieves one-shot wavefront acquisition under extreme illumination non-uniformity, reaching a dynamic range of 260 dB at a 20 Hz acquisition frequency. 
Owing to this high dynamic range and the concomitant high-intensity resolution, wavefront reconstruction errors in dim and bright sub-apertures are reduced by 59\% and 70\%, respectively, relative to conventional frame-based SHWFS. 
For dynamic wavefront sensing, the system provides kilohertz-rate centroid tracking over a megapixel field of view with microsecond-scale latency. 
Centroid localization errors are 0.18 pixels during optical alignment supervision and 0.26 pixels in high-speed turbulence observation, verifying the accuracy and reliability of the system across dynamic scenarios. 
The per-sub-aperture processing throughput reaches 420,737 Hz on a standard CPU, demonstrating high-speed real-time computation without specialized hardware acceleration. 
Together, these results establish a unified neuromorphic SHWFS framework for high-fidelity one-shot static wavefront reconstruction and real-time high-bandwidth dynamic wavefront sensing.
\end{abstract}

\setboolean{displaycopyright}{false} 

\begin{document}

\maketitle

\section{Introduction}\label{sec:intro}
Shack–Hartmann wavefront sensors (SHWFS)~\cite{platt2001history} are widely used in optical metrology~\cite{deng2025measurement,song2025dynamic,cheng2020fabrication,6155056,Wang:23}, active alignment~\cite{Schindlbeck:18,ahn2025optical,adil2017optical}, and turbulence monitoring~\cite{andrade2019estimation,wilsonSLODARMeasuringOptical2002}. 
In static metrology, the goal is high-fidelity wavefront reconstruction from spatially sampled focal-spots. 
This requires not only accurate centroid estimation for local slope recovery, but also faithful measurement of the focal-spot energy distribution to enable higher-order local approximations of the wavefront~\cite{liPhaseRetrievalUsing2014a,viegersNonlinearSplineWavefront2017,feng2018moment}.
Dynamic applications, such as active alignment and turbulence sensing, impose an even more demanding requirement: the sensor need to track rapid wavefront fluctuations with high temporal resolution, low latency, and sustained real-time throughput, often under constrained illumination and acquisition bandwidth. 
Together, these demands call for a wavefront sensing framework that can operate effectively in both static and dynamic regimes.
Conventional frame-based SHWFS, however, are constrained by two fundamental limitations rooted in CMOS/CCD imaging architectures. 
First, their limited dynamic range in intensity causes simultaneous saturation in bright sub-apertures and underexposure in dim ones, distorting the focal-spot energy distributions required for higher-order local wavefront reconstruction. 
Second, they face an inherent trade-off between spatial resolution and frame rate, making it difficult in practice to simultaneously achieve megapixel-scale spatial resolution and kilohertz-class temporal resolution, thereby constraining high-bandwidth dynamic wavefront sensing.

Event-based neuromorphic sensors~\cite{gallegoEventBasedVisionSurvey2022} provide an appealing alternative to frame-based imaging for wavefront sensing. 
By asynchronously reporting per-pixel irradiance changes with microsecond temporal resolution, they offer low-latency readout and naturally concentrate acquisition bandwidth on regions where signal is present. 
This property is particularly attractive for SHWFS, where the informative content is confined to sparse focal-spot arrays occupying only a small fraction of the sensor plane.
In practice, however, existing event-based SHWFS~\cite{grose2024convolutional,wangAngleBasedNeuromorphicWave2025} are still constrained by the passive sensing paradigm: they rely on motion-triggered events produced by focal-spot displacement alone. 
Consequently, they fail in static or slowly varying scenes, exhibit poor signal-to-noise ratio under weak dynamics, and lack a robust mechanism for distinguishing signal from background artifacts. 
As a result, the potential of neuromorphic sensing in SHWFS remains largely untapped.

\begin{figure*}[ht!]
  \centering
   \includegraphics[width=\textwidth]{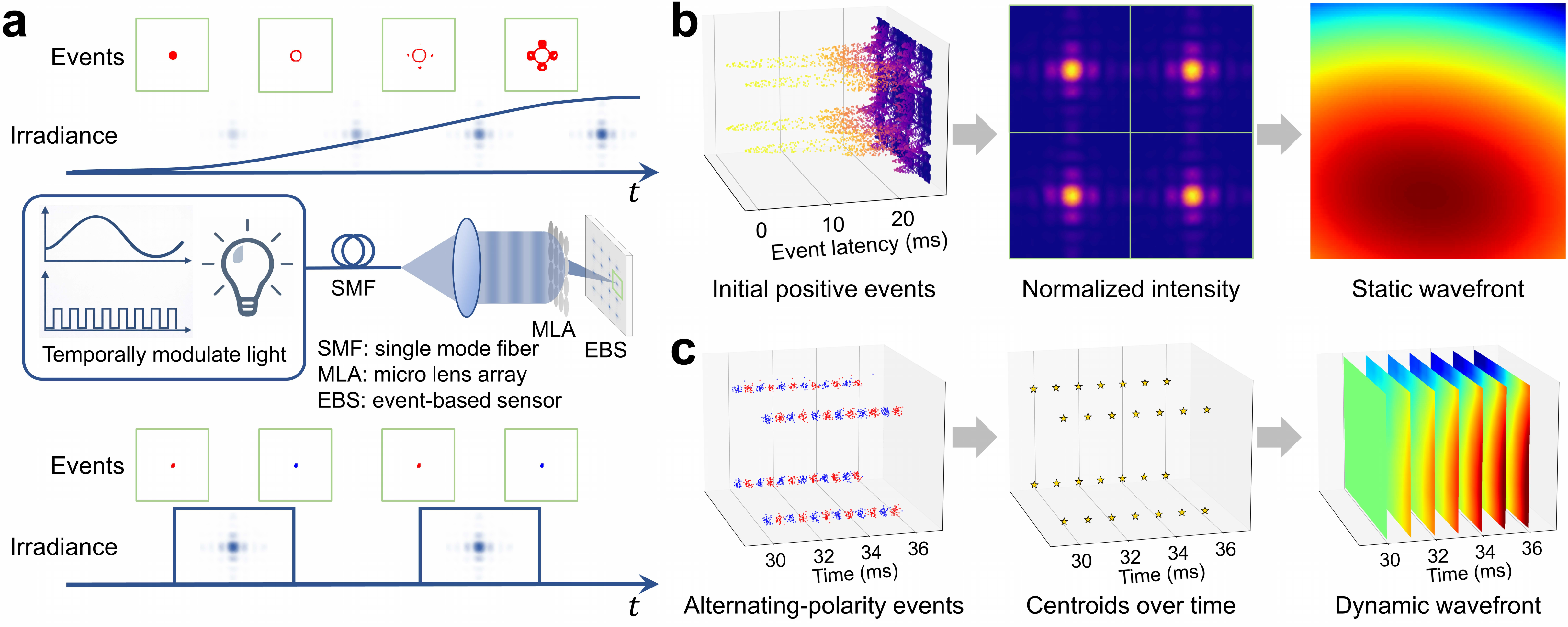}
    \caption{\textbf{Overview of the proposed framework and its dual operating modes.}
    \textbf{a} System overview: temporally modulated illumination produces periodic irradiance variations on the event-based sensor. 
    Under low-frequency sine-wave modulation (top), positive events (red dots) are triggered at different times across the focal-spot profile as the irradiance gradually rises, thereby encoding spatial intensity into event latency. 
    Under high-frequency square-wave modulation (bottom), periodic irradiance switching generates alternating-polarity positive (red) and negative (blue) event pairs within each modulation cycle.
    \textbf{b} Static wavefront sensing mode: under \qty{20}{\hertz} sine-wave modulation, the latency distribution of initial positive events is mapped to focal-spot intensity, enabling reconstruction of high-dynamic-range, high-intensity-resolution focal-spot profiles. A moment-based reconstruction method~\cite{feng2018moment} is then applied to recover higher-order local wavefront structure from the reconstructed intensity distribution.
    \textbf{c} Dynamic wavefront sensing mode: under \qty{1}{\kilo\hertz} square-wave modulation, alternating-polarity event pairs are generated in each modulation cycle for real-time centroid localization. By tracking these centroids over time, a slope-based reconstruction method recovers dynamic wavefront evolution at kilohertz rates.}
\label{fig:system_overview}
\end{figure*}

To overcome these limitations, we propose Event Temporal-modulation SHWFS (EvTem-SHWFS), a unified framework that combines neuromorphic sensing with active temporal modulation (\Cref{fig:system_overview}a). 
By periodically modulating the illumination, the system converts focal-spot intensity distributions into deterministic event streams on the sensor plane.
The same hardware platform then supports two complementary operating modes tailored to different wavefront sensing regimes.
For static wavefront sensing, we employ low-frequency sine-wave modulation (\qty{20}{\hertz}), which encodes fine-grained focal-spot intensity variations into microsecond-scale event latencies (\Cref{fig:system_overview}b). This enables high-fidelity recovery of focal-spot energy distributions and thereby supports higher-order local wavefront reconstruction~\cite{feng2018moment}.
For dynamic wavefront sensing, we use high-frequency square-wave modulation (\qty{500}{\hertz}--\qty{1}{\kilo\hertz}), which generates alternating-polarity event pairs within each focal-spot region (\Cref{fig:system_overview}c). These alternating-polarity events are selectively extracted using a Polarity Flip Filter (PFF) and then spatially localized for centroid tracking. In this mode, EvTem-SHWFS enables kilohertz-rate, real-time dynamic wavefront sensing while suppressing background activity and reducing computational load. 
Together, these two modes enable a single sensing architecture to seamlessly support both high-fidelity static reconstruction and high-bandwidth dynamic wavefront sensing.

We validate EvTem-SHWFS across three representative wavefront sensing scenarios.
\textbf{Static optical metrology:} for a wide-angle lens with severe aperture illumination non-uniformity, EvTem-SHWFS recovers reliable focal-spot intensity information from both dim and bright sub-apertures within a single acquisition, reducing wavefront reconstruction error by 59.1\% and 70.0\%, respectively, relative to frame-based SHWFS.
\textbf{Active optical alignment:} during multi-degree-of-freedom lens assembly, with wavefront sensing performed at \qty{500}{\hertz}, EvTem-SHWFS achieves a mean centroid localization error of $0.18 \pm 0.12$ pixels, compared with $2.75 \pm 2.59$ pixels for the event-based baseline EBWFNet~\cite{grose2024convolutional}, while sustaining a per-sub-aperture throughput of \qty{420737}{\hertz} on a standard CPU.
\textbf{High-speed turbulence sensing:} under \qty{1}{\kilo\hertz} operation, EvTem-SHWFS captures flame-induced transient wavefront perturbations with a centroid localization error of $0.26 \pm 0.06$ pixels, substantially lower than the $1.87 \pm 1.18$ pixels of the passive event-based baseline EBWFNet. 
These results demonstrate the effectiveness of EvTem-SHWFS across diverse scenarios, highlighting its capability to deliver both accurate static reconstruction and robust high-speed dynamic sensing within a single framework.

\begin{figure*}[ht!]
\centering
\includegraphics[width=0.8\textwidth]{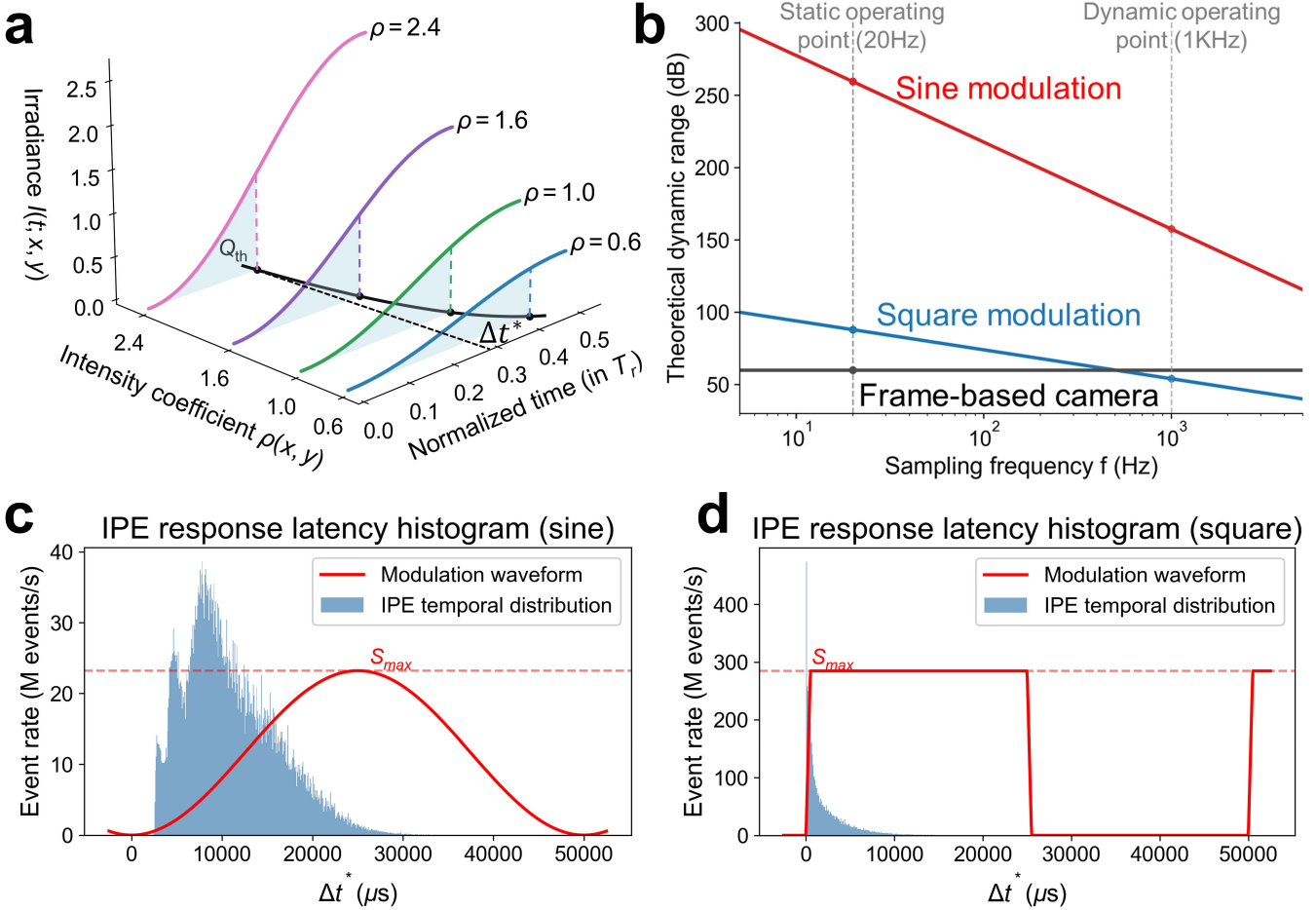}
    \caption{\textbf{Temporal encoding principle and modulation-dependent operating regimes of EvTem-SHWFS.}
    \textbf{a} Temporal encoding of focal-spot intensity under sinusoidal modulation. Different local intensities $\rho$ produce different irradiance curves $I(t;x,y)$ and trigger the initial positive event (IPE) at different latencies $\Delta t^*$ when the integrated charge reaches the threshold $Q_{th}$. The resulting $\Delta t^*$--$\rho$ mapping serves as a lookup table for pixel-wise intensity reconstruction.
    \textbf{b} Theoretical dynamic range (DR) under different modulation conditions. Sine modulation provides a $3\times$ higher log-scale DR than square modulation. At \qty{20}{\hertz}, the theoretical DR reaches $\sim$\qty{260}{\decibel}, far exceeding the fixed \qty{60}{\decibel} intensity DR of conventional frame-based cameras. At \qty{1}{\kilo\hertz}, the shorter integration window reduces the DR of square modulation below \qty{60}{\decibel}, suppressing weak side lobes and background responses.
    \textbf{c-d} Experimentally measured IPE latency histograms under sine and square modulation. Sine modulation distributes events over a broad temporal window, enabling high-resolution intensity reconstruction, whereas square modulation concentrates events near the onset of each cycle, producing a temporally compact response suitable for rapid centroid localization.}
\label{fig:principle}
\end{figure*}

\section{Principle}\label{sec:principle}

\subsection{Temporal modulation model and intensity reconstruction}

The core of EvTem-SHWFS is an active temporal encoding mechanism that maps intensity into event timing. 
As shown in \Cref{fig:principle}a, the light source is driven by a periodic modulation signal $S(t)$, illustrated here over half a sinusoidal cycle, such that the irradiance at each pixel $(x,y)$ on the sensor plane is
\begin{equation}
\label{eq:irradiance}
I(t; x, y) = \rho(x, y) S(t), \quad 0 \le t \le T_r,
\end{equation}
where $\rho(x, y)$ denotes the static intensity coefficient of the focal-spot profile to be recovered, and $T_r$ is the modulation period. 
According to the integral-to-threshold response model~\cite{bao2025temporal}, an initial positive event (IPE) is triggered with latency $\Delta t^*$ when the accumulated photon-generated charge reaches the threshold $Q_{th}$.
\begin{equation}
\label{eq:ipe_integral}
\rho(x, y) = \frac{Q_{th}}{\int_{0}^{\Delta t^*} S(t)\,\mathrm{d}t}.
\end{equation}

\Cref{eq:ipe_integral} shows that the spot intensity $\rho(x, y)$ is encoded into the IPE latency $\Delta t^*$. 
As illustrated in \Cref{fig:principle}a, the relationship between \(\rho\) and the corresponding IPE latency is projected onto the base plane as a black curve. This latency--intensity curve serves as a lookup table for pixel-wise intensity reconstruction.

\subsection{Theoretical analysis and modulation operating regimes}

The sensing behavior of EvTem-SHWFS is fundamentally governed by the modulation waveform and frequency. 
From \Cref{eq:ipe_integral}, the measurable intensity coefficient \(\rho\) is inversely proportional to the integral of \(S(t)\) over \(\Delta t^*\). 
Consequently, the modulation waveform and frequency jointly determine the achievable intensity dynamic range (DR) and the event temporal distribution.

The measurable DR is given by the ratio between the maximum and minimum detectable $\rho$, corresponding to the shortest and longest admissible integration times. 
The shortest integration time is limited by the minimum temporal resolution of the event sensor, $\tau_{\text{min}}$ (typically \qty{1}{\micro\second} for commercial devices), whereas the longest integration time is bounded by half the modulation period, $T_r/2$, for a symmetric waveform. 
Substituting these bounds into \Cref{eq:ipe_integral} yields the theoretical DR for different modulation waveforms, as shown in \Cref{fig:principle}b.

In the low-frequency regime, the long integration window enables a large measurable DR. 
In particular, sine-wave modulation achieves a \(3\times\) higher log-scale DR than square-wave modulation. 
At the static operating point (\qty{20}{\hertz}), sine modulation reaches a theoretical DR of \qty{260}{\decibel}, far exceeding the fixed \qty{60}{\decibel} intensity DR of frame-based cameras. 
A detailed derivation of the theoretical DR is provided in Supplementary Section 2. 
This extended DR allows EvTem-SHWFS to recover the full focal-spot intensity distribution in a single acquisition, spanning the bright primary lobe and the weak side lobes.
As the modulation frequency increases, the integration window \(T_r/2\) correspondingly shortens, driving the system into a high-frequency regime with reduced measurable DR. 
At the dynamic operating point (\qty{1}{\kilo\hertz}), the maximum integration window shrinks to \qty{0.5}{\milli\second}, and the theoretical DR of square-wave modulation drops below \qty{60}{\decibel} (\Cref{fig:principle}b). 
Under this short integration window, only high-intensity regions of the focal spot, such as the primary lobe, can accumulate sufficient charge to trigger events, whereas weak side lobes and background are inherently suppressed. 
This effect is directly reflected in \Cref{fig:system_overview}c, where \qty{1}{\kilo\hertz} square-wave modulation produces alternating positive and negative event pairs primarily from the primary-lobe region.
The experimentally measured IPE latency histograms at the same \qty{20}{\hertz} modulation frequency further highlight the waveform-dependent temporal responses of EvTem-SHWFS . 
Under sine-wave modulation, the IPE latencies are distributed over a broad temporal window (\Cref{fig:principle}c), providing fine temporal quantization for high-resolution intensity reconstruction. 
Under square-wave modulation, by contrast, the histogram exhibits a sharp peak near the onset of each cycle (\Cref{fig:principle}d), indicating a temporally concentrated response that is more favorable for rapid centroid localization.

In practice, these observations provide a simple modulation-design guideline. 
For static or slowly varying wavefronts, low-frequency sine-wave modulation is preferred because it maximizes DR and preserves detailed focal-spot intensity distributions for higher-order local wavefront reconstruction~\cite{feng2018moment}. 
For dynamic wavefront sensing, high-frequency square-wave modulation is preferred because it produces sparse, temporally concentrated event responses that are well suited to rapid and real-time centroid localization for slope-based reconstruction. 
Thus, by tuning the modulation waveform and frequency, EvTem-SHWFS can be configured to operate continuously between high-fidelity intensity reconstruction and high-speed centroid-based sensing within a single hardware architecture.

\section{Results}
\label{sec:results}
We evaluate EvTem-SHWFS across three representative wavefront sensing scenarios, each corresponding to a distinct operating regime of the temporal modulation framework. 
The selected scenarios cover key application domains of Shack–Hartmann wavefront sensing: static optical metrology, active optical alignment, and high-speed turbulence sensing. 
Together, these experiments show how EvTem-SHWFS adapts to diverse sensing requirements—from high-fidelity recovery of focal-spot intensities to kilohertz-rate tracking of dynamic wavefront evolution—within a unified sensing architecture.

\begin{figure*}[ht!]
  \centering
   \includegraphics[width=0.75\textwidth]{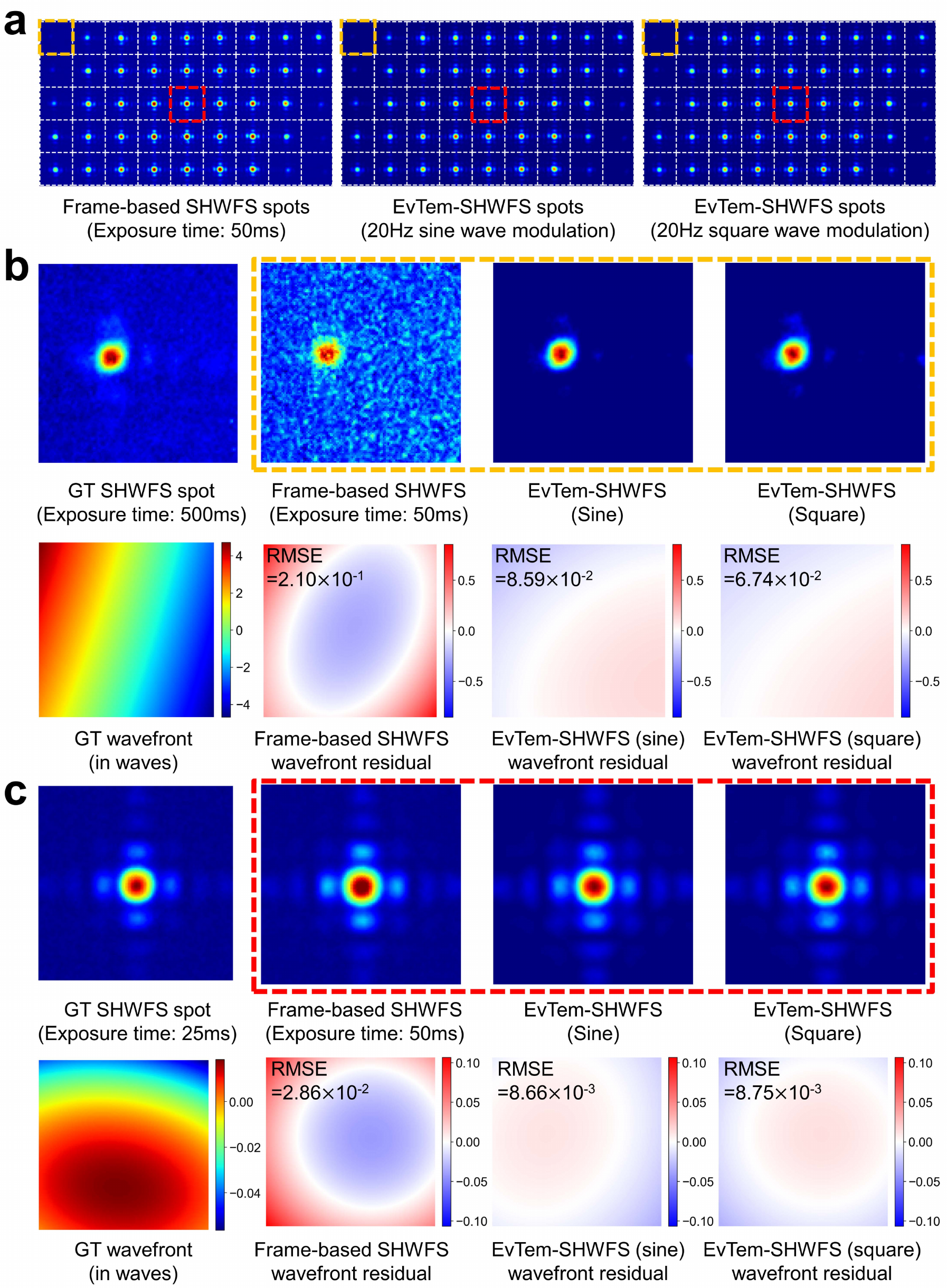}
    \caption{\textbf{Quantitative performance comparison of static wavefront sensing.} 
    \textbf{a} Full-field spot images: frame-based SHWFS (\qty{50}{\milli\second}) vs. EvTem-SHWFS (\qty{20}{\hertz} sine/square). 
    \textbf{b} Peripheral sub-aperture (yellow box) and \textbf{c} central sub-aperture (red box) images: from left to right—ground truth (GT), frame-based (\qty{50}{\milli\second}), EvTem-SHWFS (sine), and EvTem-SHWFS (square). Corresponding GT wavefronts and wavefront residuals are shown below.}\label{fig:static}
\end{figure*}

\subsection{Static wavefront sensing under non-uniform aperture illumination}

Wide-angle lens metrology presents a challenging static wavefront sensing scenario because the illumination across the aperture can vary by orders of magnitude, from a bright center to a dim periphery~\cite{song2025dynamic}. 
Under a single exposure, conventional frame-based SHWFS therefore suffers from simultaneous saturation in central sub-apertures and underexposure in peripheral ones. 
This directly distorts the focal-spot energy distributions required for moment-based wavefront reconstruction~\cite{feng2018moment}, which relies on accurate first-order moments for local slope estimation and second-order central moments for local curvature estimation. 
Although multiple exposures can partially alleviate this problem, they come at the cost of measurement efficiency. 
In contrast, EvTem-SHWFS uses its high-dynamic-range static operating regime to recover complete focal-spot intensity distributions across the aperture within a single acquisition.

We perform a quantitative comparison on the inverse-projected exit wavefront of a \qty{4}{\milli\meter} lens. 
A \qty{620}{\nano\meter} LED coupled to a single-mode fiber (\qty{3.5}{\micro\meter} core diameter) serves as the light source, and is driven at \qty{20}{\hertz} with either sine-wave or square-wave modulation. 
A 50\% beam splitter placed after the MLA simultaneously directs the focal-spots to an event-based sensor and a frame-based camera. 
The two cameras are spatially calibrated with a reprojection error of 0.14 pixels, and the frame-based exposure is synchronized to the modulation signal.

The results are summarized in \Cref{fig:static}. 
\Cref{fig:static}a compares the full-aperture spot images. Owing to its extended intensity DR, EvTem-SHWFS captures the full aperture without saturation or underexposure, whereas the frame-based SHWFS (\qty{50}{\milli\second} exposure) suffers from both under the limited \qty{60}{\decibel} DR. 
To highlight the impact of DR limitation on wavefront reconstruction, we further examine two representative sub-apertures: a dim peripheral region (\Cref{fig:static}b) and a bright central region (\Cref{fig:static}c).
We first consider the dim peripheral case. 
In \Cref{fig:static}b, a \qty{500}{\milli\second} frame-based capture is used as the ground truth (GT) for quantitative comparison. 
At the practical exposure time of \qty{50}{\milli\second}, the frame-based spot is nearly buried in background noise, severely degrading estimation of both the first-order and second-order moments of the focal-spot intensity profile.
By contrast, EvTem-SHWFS preserves the focal-spot energy distribution even in these low-light regions. 
In particular, square-wave modulation exhibits higher sensitivity under weak illumination because its near-instantaneous triggering mechanism better preserves subtle diffraction side lobes. 
After adaptive thresholding and moment-based wavefront reconstruction~\cite{feng2018moment}, the wavefront RMSE for the peripheral sub-aperture is \(2.1 \times 10^{-1}\) waves for frame-based SHWFS, versus \(8.59 \times 10^{-2}\) waves for EvTem-SHWFS with sine modulation (59\% reduction) and \(6.74 \times 10^{-2}\) waves with square modulation (68\% reduction).
We next consider the bright central case in \Cref{fig:static}c. 
Here, saturation in the frame-based image clips the spot core, distorts the point spread function morphology, and destroys the second-order moment information required for local curvature sensing. 
EvTem-SHWFS, in contrast, preserves the full intensity gradient from the bright primary lobe to the weak surrounding structures without clipping. 
As a result, the reconstructed local wavefront achieves an RMSE of \(8.66 \times 10^{-3}\) waves with sine modulation, compared with \(2.86 \times 10^{-2}\) waves for frame-based SHWFS, reducing the reconstruction error by about 70\%.

These results show that, in the static operating regime, EvTem-SHWFS enables one-shot wavefront sensing under strongly non-uniform aperture illumination, while preserving the focal-spot intensity distributions required for higher-order local wavefront reconstruction. 
It therefore overcomes the intensity dynamic range bottleneck that limits frame-based SHWFS in static metrology.

\begin{figure*}[ht!]
  \centering
   \includegraphics[width=0.8\textwidth]{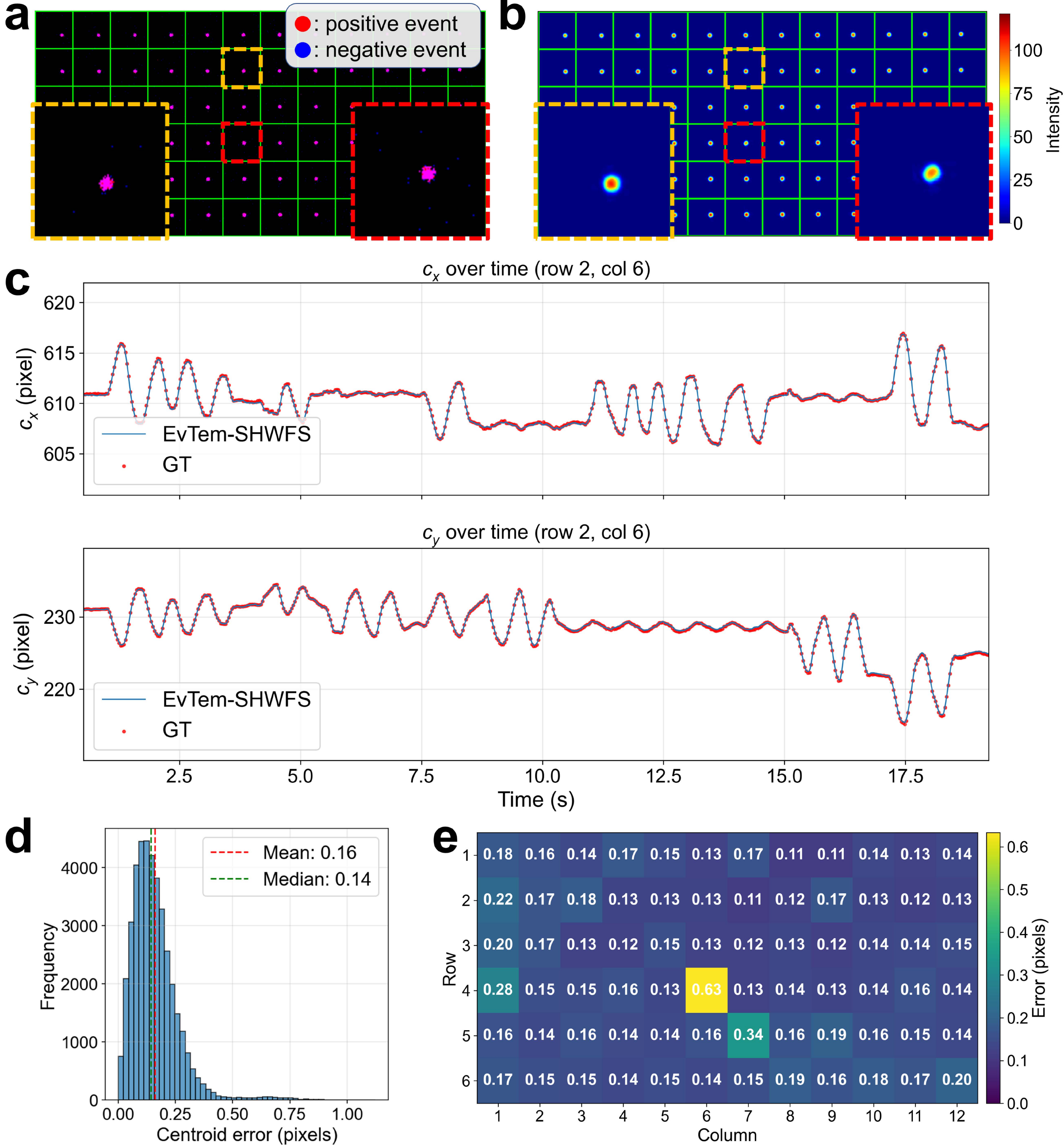}
    \caption{\textbf{Real-time dynamic wavefront sensing at \qty{500}{\hertz}.}
    \textbf{a} Accumulated event frame over a \qty{2}{\milli\second} window (starting at $t = 0.52$~s). Red and blue dots represent positive and negative events; magenta clusters indicate alternating-polarity event pairs.
    \textbf{b} Corresponding conventional intensity frame (\qty{20}{\milli\second} exposure, 30~FPS) in pseudo-color, used to derive the proxy ground-truth centroids.
    \textbf{c} Centroid trajectories ($c_x, c_y$) for a representative sub-aperture (row 2, col 6): EvTem-SHWFS at \qty{500}{\hertz} (blue line) and GT at \qty{30}{\hertz} (red dots).
    \textbf{d} Histogram of centroid localization errors across all sub-apertures at GT timestamps. Mean error: 0.16 pixels; median error: 0.14 pixels.
    \textbf{e} Spatial error distribution map. Sub-aperture (4,6) shows an elevated error of 0.63 pixels.}
    \label{fig:AA}
\end{figure*}

\subsection{Real-time dynamic wavefront sensing for optical alignment}

In active optical alignment, wavefront sensing can in principle provide direct feedback for correcting aberrations induced by component displacement~\cite{Schindlbeck:18,ahn2025optical,adil2017optical}. 
In practice, however, most alignment processes remain largely open-loop, relying on static measurements or low-rate estimation combined with predictive models~\cite{liuApplicationDeepLearning2024}. 
This paradigm is insufficient for transient wavefront variations arising during multi-degree-of-freedom assembly. 
Enabling closed-loop alignment instead requires a wavefront sensor that can simultaneously capture rapid aberration dynamics with high spatiotemporal resolution and process the resulting data in real time on standard hardware. 
Frame-based SHWFS fall short on both requirements because of the trade-off between spatial and temporal resolution~\cite{Wu:23}, while kilohertz-rate full-aperture centroid localization typically depends on specialized hardware such as FPGAs~\cite{ye2025real} or GPUs~\cite{Yang:25}. 
In this section, we show that EvTem-SHWFS overcomes these two barriers, providing both high-speed wavefront tracking and real-time processing capability for closed-loop active optical alignment.

We monitor the dynamic evolution of the exit wavefront during lens assembly using the same hardware configuration as in the static experiments, except that the illumination is driven by a \qty{500}{\hertz} square-wave modulation for dynamic sensing. 
A frame-based camera (\qty{30}{\hertz}, \qty{20}{\milli\second} exposure) is placed at the secondary output of a beam splitter after the MLA and spatially registered to the event camera; the centroids extracted from these intensity frames serve as proxy ground truth (GT). 
The \(1280 \times 720\)-pixel sensor is partitioned into a \(14 \times 8\) sub-aperture array, among which the central \(12 \times 6\) sub-apertures are used for quantitative evaluation, as focal-spots near the boundary may drift outside the sensor area during sensing.
To benchmark against passive event-based sensing, we repeat the same experiment with active modulation disabled, so that the event camera records only motion-triggered events.
These passive event streams are then processed using EBWFNet~\cite{grose2024convolutional}, enabling direct comparison with EvTem-SHWFS in both centroid accuracy and computational throughput.

We first examine a representative 18-second sequence recorded during the assembly of a \qty{35}{\milli\meter} lens, as summarized in \Cref{fig:AA}. 
\Cref{fig:AA}a shows the accumulated event frame over a \qty{2}{\milli\second} window at \(t = 0.52\)~s. 
Within each sub-aperture, a compact magenta cluster marks the alternating-polarity event pair, consistent with the spot locations observed in the corresponding frame-based image in \Cref{fig:AA}b.
The temporal tracking capability of EvTem-SHWFS is illustrated in \Cref{fig:AA}c for the representative sub-aperture indicated by the yellow box (row 2, column 6). 
The \qty{500}{\hertz} centroid trajectory recovered by EvTem-SHWFS closely follows the low-frequency trend of the \qty{30}{\hertz} proxy ground truth, while additionally resolving irregular sub-millisecond fluctuations between successive GT samples. 
These intermediate oscillations reveal transient wavefront variations that are not observable with the frame-based sensor.
The centroid localization accuracy is quantified in \Cref{fig:AA}d. 
Across all evaluated sub-apertures and the full sequence, EvTem-SHWFS achieves a mean centroid error of 0.16 pixels and a median error of 0.14 pixels. 
Notably, the mean error is close to the 0.14-pixel reprojection error of the spatial calibration between the event-based and frame-based cameras, indicating that the dynamic tracking precision is approaching the calibration limit of the experimental setup.
The spatial distribution of centroid errors is shown in \Cref{fig:AA}e. 
One outlier appears at sub-aperture (4,6), where the error rises to 0.63 pixels. 
Inspection of the corresponding focal-spot in \Cref{fig:AA}b reveals a persistent local distortion caused by a microlens defect, which introduces a fixed bias between the tracked centroid and the proxy ground truth. 
A detailed analysis of this outlier is provided in Supplementary Section 5.
Aside from this defect-affected region, centroid errors remain low across the array, with slightly higher errors near the aperture boundary, consistent with the lower spot intensity and reduced signal-to-noise ratio in these regions.
In addition to accurate tracking, EvTem-SHWFS also operates comfortably in real time on standard CPU hardware. 
For this representative 18-second sequence, the PFF-based centroid extraction pipeline achieves a per-sub-aperture throughput of approximately \qty{417000}{\hertz}, corresponding to a full-aperture throughput of \qty{3585}{\hertz} for the \(14 \times 8\) sub-aperture array. 
This exceeds the \qty{500}{\hertz} sensing rate by more than a factor of seven, confirming that the incoming event stream can be processed online without backlog.

\begin{table*}[htbp]
\centering
\caption{Centroid localization error (in pixels) across different assembly scenarios.}
\label{tab:accuracy_detail}
\begin{tabular}{cccccc}
\toprule
Lens & Assembly DOF & \makecell{EvTem-SHWFS \\ (Ours)} & \makecell{EBWFNet~\cite{grose2024convolutional} \\ (Public)} & \makecell{EBWFNet \\ (Fine-tuned)} \\
\midrule
\multirow{4}{*}{24 mm, F/1.5} & Axial spacing          & $0.29 \pm 0.15$ & $6.12\pm 2.06$ & $3.47\pm 1.96$  \\
                       & Sagittal tilt    & $0.27 \pm 0.13$ & $6.11\pm 1.47$ & $2.04 \pm 1.32$  \\
                       & Meridional tilt  & $0.13 \pm 0.10$ & $4.76\pm 1.65$ & $2.05\pm 1.21$  \\
                       & Combined tilt    & $0.14 \pm 0.08$ & $5.04\pm 2.13$ & $1.42\pm 0.78$  \\
\midrule
\multirow{4}{*}{35 mm, F/1.5} & Axial spacing         & $0.29\pm 0.14$ & $12.00\pm 7.66$ & $3.17\pm 1.59$  \\
                       & Sagittal tilt    & $0.12\pm 0.07$ & $4.97\pm 5.48$ & $2.16 \pm 1.98$  \\
                       & Meridional tilt  & $0.12\pm 0.09$ & $4.46\pm 4.47$ & $4.76\pm 4.28$  \\
                       & Combined   tilt  & $0.13\pm 0.09$ & $4.49\pm 2.47$ & $2.00\pm 1.34$  \\
\midrule
\multirow{4}{*}{50 mm, F/1.6} & Axial spacing           & $0.24\pm 0.13$ &  $13.97\pm 7.34$  & $3.08\pm 2.12$   \\
                       & Sagittal tilt    & $0.16\pm 0.09$ & $4.87\pm 5.04$ & $3.53\pm 4.69$  \\
                       & Meridional tilt  & $0.17\pm 0.09$ & $3.51\pm 1.98$ & $3.05\pm 3.17$  \\
                       & Combined   tilt  & $0.19\pm 0.12$ & $5.64\pm 4.94$ & $2.85\pm 2.10$  \\
\midrule
\multicolumn{2}{c}{Overall} 
& $0.18 \pm 0.12$ 
& $6.31 \pm 5.27$ 
& $2.75 \pm 2.59$ \\
\bottomrule
\end{tabular}
\end{table*}

We next extend the evaluation to a broader dataset covering three lenses (24 mm, 35 mm, and 50 mm) and four assembly operations (axial spacing, sagittal tilt, meridional tilt, and combined tilt). 
As summarized in \Cref{tab:accuracy_detail}, EvTem-SHWFS achieves an overall mean centroid localization error of \(0.18 \pm 0.12\) pixels, substantially lower than both the public EBWFNet baseline (\(6.31 \pm 5.27\) pixels) and its fine-tuned version (\(2.75 \pm 2.59\) pixels).
For reference, the public EBWFNet weights are trained on its original public dataset, collected with an EVK3 sensor and characterized by relatively large spot motions, whereas the fine-tuned version is retrained on a training set collected from our active alignment setup acquired with an EVK4 sensor. 
Because EBWFNet expects \(50 \times 50\) pixel sub-apertures, our \(96 \times 96\) pixel event patches are downsampled before inference and the predicted centroid coordinates are then rescaled accordingly. This same preprocessing is applied to both the public and fine-tuned models.
Several trends are apparent from \Cref{tab:accuracy_detail}. 
First, axial spacing is consistently more challenging than the other assembly degrees of freedom for all three methods. 
Unlike tilt perturbations, axial spacing modifies not only focal-spot positions but also their energy distributions: as the wavefront becomes more convex, central sub-apertures brighten while peripheral ones dim. 
For EvTem-SHWFS, the reduced spot intensity lowers the signal-to-noise ratio of the alternating-polarity event pairs and thus degrades centroid precision. For EBWFNet, intensity-induced events interfere with the motion-triggered events on which the passive method relies, leading to the same trend.
Second, the public EBWFNet model exhibits particularly large errors for axial spacing on the 35 mm and 50 mm lenses. 
For these longer-focal-length lenses, axial spacing produces smaller spot displacements, especially in central sub-apertures, resulting in much sparser event streams than those seen in the public training set. This mismatch leads to poor generalization. After fine-tuning on our dataset, the errors decrease substantially, returning to levels comparable to those for the 24 mm lens.
In contrast, EvTem-SHWFS maintains consistently low errors across all lenses and assembly operations, with standard deviations that remain much smaller than those of either EBWFNet baseline. 
These results indicate that the proposed active event-generation mechanism not only improves centroid accuracy, but also provides substantially stronger robustness and cross-scenario generalization than passive event-based sensing.

\begin{table*}[htbp]
\centering
\begin{threeparttable}
\caption{Per-sub-aperture processing time and throughput.}
\label{tab:throughput_breakdown}
\begin{tabular}{lcccc}
\toprule
Method & Preprocessing (\textmu s) & Centroid localization (\textmu s) & Throughput (Hz) \\
\midrule
EvTem-SHWFS (CPU) & $1.16\pm 0.31$ & $1.22\pm 0.20$  & $420{,}737$ \\
EBWFNet (CPU) & $(9.66 \pm 3.55)\times 10^3$ &  $(1.05 \pm 0.12)\times 10^3$  & $93$  \\
EBWFNet (CPU+GPU) & $(10.41 \pm 3.78)\times 10^3$ &  $(0.18 \pm 0.07)\times 10^3$  & $94$  \\
\bottomrule
\end{tabular}
\begin{tablenotes}
\small
\item CPU: Intel i9-13900KF; GPU: NVIDIA RTX 4090.
\end{tablenotes}
\end{threeparttable}
\end{table*}

Finally, we evaluate whether EvTem-SHWFS can satisfy the computational requirements of real-time closed-loop alignment. 
As summarized in \Cref{tab:throughput_breakdown}, we compare the average processing time per sub-aperture sample, including both preprocessing and centroid localization, and define the throughput as the reciprocal of their sum.
For EvTem-SHWFS, preprocessing consists of applying the Polarity Flip Filter (PFF) to the raw event stream, followed by direct centroid localization from the extracted polarity event pairs. 
The entire pipeline runs on a CPU.
For EBWFNet, preprocessing is also performed on the CPU and includes the construction and packing of time-ordered recent event (TORE) volumes for network input. 
During inference, the TORE volumes from all 72 evaluated sub-apertures are batched together, and centroid localization is then executed on either a CPU or a GPU.
As shown in \Cref{tab:throughput_breakdown}, EvTem-SHWFS achieves a per-sub-aperture throughput of \qty{420737}{\hertz} on a standard CPU. 
In contrast, EBWFNet reaches only \qty{93}{\hertz} on CPU and \qty{94}{\hertz} on CPU+GPU.
For EBWFNet, preprocessing dominates the total latency, requiring \qty{9.66}{\milli\second} per sub-aperture on CPU, whereas GPU acceleration reduces only the inference time and therefore provides negligible improvement in end-to-end throughput. 
This identifies event-to-volume conversion, rather than network inference, as the primary bottleneck of the passive event-based pipeline.
By contrast, EvTem-SHWFS operates directly on the asynchronous event stream without explicit frame construction or neural-network inference. 
Its PFF-based centroid extraction therefore preserves the event-native processing paradigm and achieves high computational efficiency on standard hardware. 
These results show that EvTem-SHWFS is not only more accurate than EBWFNet, but also substantially better suited to the real-time demands of closed-loop active optical alignment.
A complexity analysis of the EvTem-SHWFS centroid localization algorithm is provided in Supplementary Section 4.

Taken together, these results demonstrate that EvTem-SHWFS enables accurate, high-speed wavefront tracking during active optical alignment while maintaining real-time processing capability on standard hardware. 
By combining deterministic event generation with event-native processing, the proposed framework overcomes both the accuracy limitations of passive event-based methods and the computational bottlenecks of frame-based or volume-based pipelines. 
This makes EvTem-SHWFS particularly well suited for closed-loop alignment scenarios requiring simultaneous high spatiotemporal resolution and low-latency feedback.

\subsection{High-speed wavefront sensing for turbulence monitoring}
\begin{figure*}[ht!]
  \centering
   \includegraphics[width=0.99\textwidth]{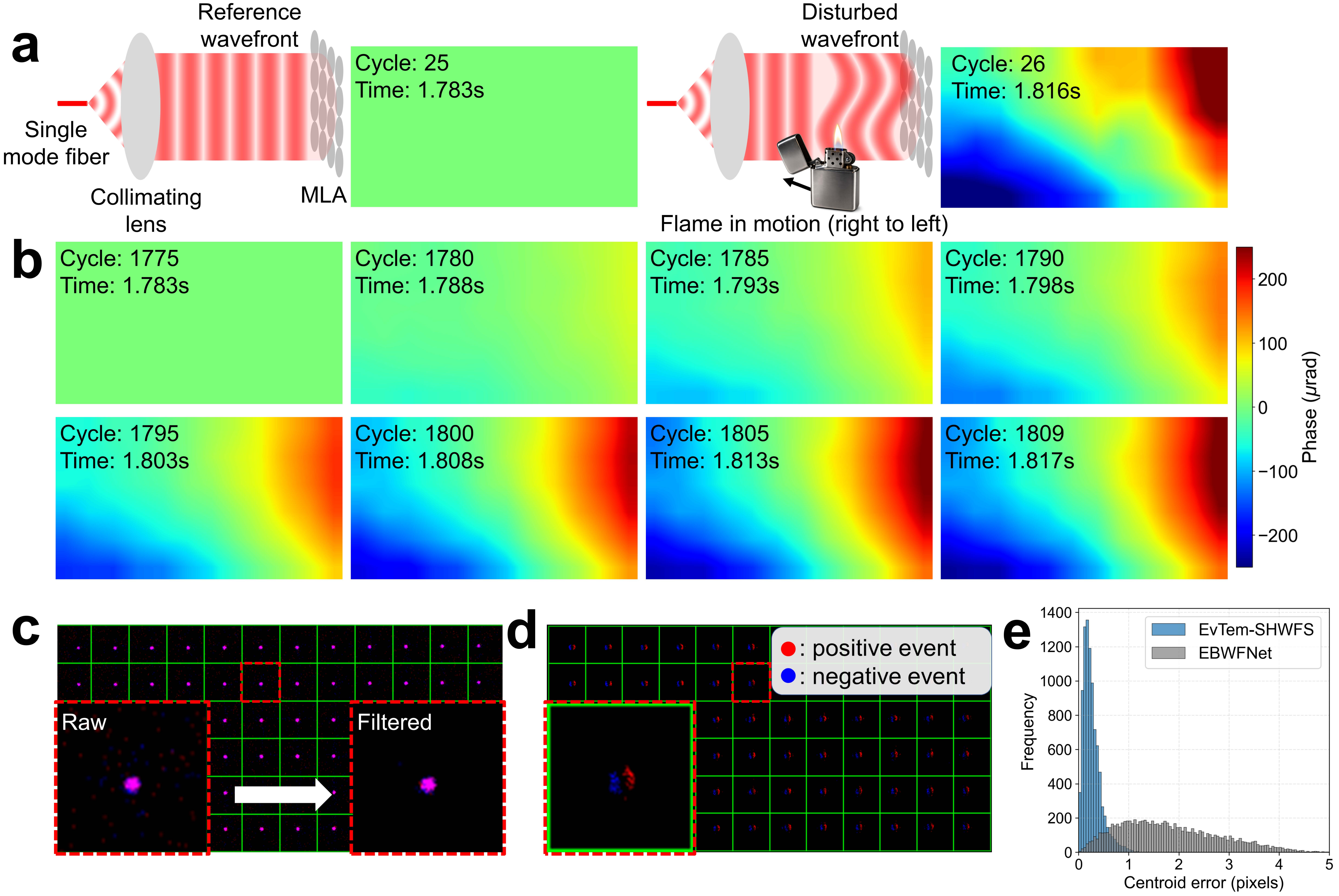}
    \caption{\textbf{Kilohertz capture of transient wavefront distortions induced by high-speed flame turbulence.} 
    \textbf{a} Experimental schematic and frame-based SHWFS results (\qty{30}{\hertz}). The flame moves from right to left; the frame-based SHWFS captures a reference state (\(t = 1.783\)~s) and a distorted state (\(t = 1.816\)~s).
    \textbf{b} EvTem-SHWFS wavefront snapshots acquired at \qty{1}{\kilo\hertz} and displayed at \qty{5}{\milli\second} intervals.
    \textbf{c} Event frame under \qty{1}{\kilo\hertz} modulation (\qty{1}{\milli\second} window). The magnified view of sub-aperture (2,6) shows the raw event frame (left) and the same frame after applying the Polarity Flip Filter (right). Background events (red and blue) are effectively suppressed, while the alternating-polarity focal-spot event pairs (magenta) are preserved.
    \textbf{d} Event frame under constant illumination (\qty{10}{\milli\second} window). Positive (red) and negative (blue) event clusters, generated by spot motion, form spatially separated structures that roughly outline the focal-spot boundaries.
    \textbf{e} Centroid error histogram. EvTem-SHWFS (blue): \(0.26 \pm 0.06\) pixels; EBWFNet~\cite{grose2024convolutional} (gray): \(1.87 \pm 1.18\) pixels.}
\label{fig:flame_turbulence}
\end{figure*}

Flow- or heat-induced optical turbulence gives rise to rapid refractive-index fluctuations~\cite{owens1967optical}, which in turn produce dynamic wavefront distortions over a broad temporal bandwidth~\cite{wang2012physics,jumper2017physics,gordeyev2014experimental}. 
Accurately resolving such transient wavefront evolution is important for applications including flow diagnostics, combustion analysis, and adaptive optics~\cite{dave2024wavefront,wu2019wish}. 
However, frame-based SHWFS is fundamentally limited by its frame rate, which undersamples these fast dynamics and therefore captures only discrete wavefront states rather than their continuous temporal evolution.
In this section, we show that EvTem-SHWFS enables high-bandwidth wavefront sensing at kilohertz rates. 
Using a flame-induced turbulence field as a representative high-speed refractive-index perturbation, we demonstrate continuous tracking of dynamic wavefront evolution that is inaccessible to frame-based sensors.

The experimental configuration is largely identical to that used in the previous measurements, with all data acquired using the same megapixel-scale Prophesee EVK4 event-based sensor. 
In this experiment, the illumination is modulated by a \qty{1}{\kilo\hertz} square wave to enable millisecond-scale wavefront sensing. 
At this higher modulation frequency, the available integration time within each cycle is reduced, which decreases the effective photon flux accumulated during each sensing interval. This reduced photon budget lowers the signal-to-noise ratio and consequently makes robust wavefront sensing more challenging than in the lower-frequency operating regimes.
Under this \qty{1}{\kilo\hertz} sensing condition, we investigate transient wavefront disturbances generated by a high-speed burst flame. 
As illustrated in \Cref{fig:flame_turbulence}a, light emitted from a single-mode fiber is collimated into a reference plane wave in the absence of the flame, and the MLA focuses this wavefront onto the sensor plane to form a regular focal-spot array. 
When the flame is introduced upstream of the MLA, strong temperature gradients induce rapid air-density fluctuations, producing a dynamically distorted wavefront with complex phase variations.


\Cref{fig:flame_turbulence}a compares two successive wavefront measurements (Cycle 25 and 26) acquired by a frame-based SHWFS operating at \qty{30}{\hertz}. 
In the absence of sufficient temporal sampling, the disturbance can only be observed as two discrete states: a reference wavefront before the flame enters the optical path and a distorted wavefront after the perturbation develops. The transient evolution between these two states remains unresolved.
By contrast, EvTem-SHWFS resolves the intermediate wavefront dynamics at \qty{1}{\kilo\hertz}, as shown in \Cref{fig:flame_turbulence}b. 
For visual clarity, only every fifth reconstruction is displayed, corresponding to \qty{5}{\milli\second} intervals, while the full \qty{1}{\kilo\hertz} sequence is provided in the Supplementary Video. 
As the flame moves from right to left into the beam path, wavefront perturbations first emerge near \(t = 1.793\)~s and then grow continuously until the distorted pattern closely matches the frame-based measurement at \(t = 1.816\)~s. 
This progression reveals both the spatial structure and the temporal buildup of the flame-induced aberration, which are inaccessible to the frame-based sensor.

We next compare the raw event patterns generated by the proposed active sensing scheme and the passive motion-event baseline. 
In \Cref{fig:flame_turbulence}c, EvTem-SHWFS produces alternating-polarity event pairs within each sub-aperture under \qty{1}{\kilo\hertz} square-wave modulation. 
Although the reduced illumination at this operating point introduces additional background activity, the Polarity Flip Filter effectively suppresses events that do not follow the modulation-induced polarity alternation, while preserving the focal-spot responses. 
In contrast, under constant illumination, the passive event stream shown in \Cref{fig:flame_turbulence}d consists of spatially separated positive and negative clusters generated only by focal-spot motion. 
These motion-triggered events provide a much weaker and less structured signal for centroid localization under the present low-amplitude, high-speed turbulence condition.
This difference in signal formation is directly reflected in centroid localization accuracy, as quantified in \Cref{fig:flame_turbulence}e.
EvTem-SHWFS achieves a mean centroid error of \(0.26 \pm 0.06\) pixels, whereas the passive event-based baseline yields \(1.87 \pm 1.18\) pixels. 
Thus, even under the more challenging \qty{1}{\kilo\hertz} operating condition with reduced photon flux, EvTem-SHWFS maintains robust centroid localization and enables high-bandwidth wavefront sensing of transient turbulence dynamics.

Although real-time processing is not the primary requirement in this experiment, EvTem-SHWFS still maintains online capability. 
Owing to the higher event rate at \qty{1}{\kilo\hertz}, the per-sub-aperture throughput decreases to approximately \qty{221223}{\hertz}, corresponding to a full-aperture throughput of \qty{1975}{\hertz} for the \(14 \times 8\) sub-aperture array. 
This remains above the sensing rate, confirming that the proposed pipeline can process the incoming event stream in real time.

\section{Discussion}

EvTem-SHWFS establishes a unified framework for wavefront sensing by combining active temporal modulation with neuromorphic event sensing across both static and dynamic operating regimes. 
While the results demonstrate the promise of this approach, several limitations and opportunities for future improvement should be noted.
First, the introduction of active temporal modulation increases the volume of raw event data compared to passive event-based SHWFS approaches. 
Although the proposed PFF-based centroid localization pipeline can process this high-rate event stream in real time, the resulting data contain substantial redundancy from a storage perspective. 
A promising direction for future work is to combine modulation-induced events with motion-triggered events, enabling more efficient event utilization while retaining the robustness of the proposed sensing paradigm.
Second, the current experimental setup is constrained by the limited optical power of the light source. 
In our implementation, an LED with a maximum drive current of \qty{1}{\ampere} is coupled into a single-mode fiber with a \qty{3.5}{\micro\meter} core diameter, resulting in limited optical throughput and a low-photon-flux sensing regime.
This limitation is particularly evident under high-frequency operation: as shown in \Cref{fig:AA}a and \Cref{fig:flame_turbulence}c, the alternating-polarity event pairs in peripheral sub-apertures are visibly smaller than those in central sub-apertures, indicating that the lower effective illumination causes more severe truncation of the focal-spot profile.
Replacing the LED with a higher-radiance source, such as a laser, would enable both higher signal-to-noise ratios in the static regime and potentially extend the operating bandwidth beyond the kilohertz range.
Third, the current implementation relies on direct illumination modulation, which naturally restricts the method to active sensing scenarios. 
Exploring alternative forms of temporal modulation within the optical system itself—for example, by introducing modulation at a plane conjugate to the light source—may broaden the applicability of the approach while preserving its underlying temporal encoding mechanism.

Looking forward, we believe that EvTem-SHWFS is particularly well suited to closed-loop active optical alignment, where its combination of high temporal bandwidth and real-time processing can provide low-latency feedback for dynamic system correction. 
Beyond alignment, the ability to capture continuous wavefront evolution at kilohertz rates opens new opportunities for investigating fast physical phenomena, including turbulent flows and other transient refractive-index dynamics.
Overall, this work establishes a new paradigm for wavefront sensing based on temporal modulation and event-driven acquisition. 
We hope it will stimulate further exploration of time-domain encoding strategies and their integration with neuromorphic sensing in optical metrology and beyond.

\section{Method}\label{sec:method}
\subsection{Optical and hardware implementation}
EvTem-SHWFS is implemented based on a conventional Shack--Hartmann wavefront sensing architecture, augmented with active temporal illumination modulation and an event-based sensor.
The optical system consists of a micro-lens array (MLA; Thorlabs MLA300-14AR) and a relay lens pair (Thorlabs MAP105075-A) mounted in a lens tube assembly, followed by a 50:50 beam splitter prism (JCOPTIX BS2555-T2). The two output ports of the beam splitter are coupled to an event-based sensor (Prophesee EVK4) and a frame-based sensor (MindVision 134GC), respectively.

The illumination source is a fiber-coupled LED (JCOPTIX LEFC620), coupled into a single-mode fiber (JCOPTIX FCS1-PC/SMA-600-770) with a \qty{3.5}{\micro\meter} core diameter and a numerical aperture of 0.13. 
The fiber is mounted by a fiber flange and positioned at the rear focal point of the zoom lens under test (Edmund Optics \#29-308), such that the emerging beam probes the exit wavefront of the optical system. 
The transmitted wavefront is sampled by the MLA to form a focal-spot array on the sensor plane. 
Temporal modulation is provided by an LED driver (JCOPTIX LEC1-B), which drives the source with programmable waveforms over a frequency range from \qty{1}{\hertz} to \qty{1}{\kilo\hertz}. Low-frequency sine-wave modulation is used in the static operating regime, whereas high-frequency square-wave modulation is used in the dynamic operating regime.

The LED driver, event-based sensor, and frame-based sensor are synchronized by a common signal source. 
In the low-frequency operating regime, the three devices share a single trigger channel. The start of each modulation cycle is marked by a rising-edge signal sent to the event-based sensor, and the same edge also triggers the frame-based sensor exposure, thereby ensuring temporal alignment among illumination modulation, event acquisition, and frame acquisition. 
In the high-frequency operating regime, the LED driver is driven by a dedicated high-frequency channel, whereas the event-based sensor and frame-based sensor share a separate low-frequency trigger channel at \qty{30}{\hertz}. 
This arrangement ensures temporal alignment between event and frame measurements during quantitative comparison, while allowing the illumination to operate at kilohertz modulation frequencies.
Additional hardware implementation details, including the registration between the event-based and frame-based sensors, are provided in Supplementary Section 3.

\subsection{Streaming event filtering and centroid localization}

To extract modulation-induced events from the raw event stream, we employ a Polarity Flip Filter (PFF) that processes events in a streaming manner.
Under periodic illumination at modulation frequency \(f\), pixels exposed to the modulated signal generate alternating ON/OFF events. 
This alternating-polarity behavior provides a distinctive temporal signature of modulation-induced responses, allowing them to be distinguished from background noise and motion-induced artifacts.

For each pixel, the filter monitors its event sequence \(\{(t_i, p_i)\}\), where \(t_i\) and \(p_i \in \{+1,-1\}\) denote the timestamp and polarity of the \(i\)-th event, respectively. An event is accepted as a valid modulation-induced event if it satisfies
\begin{equation}
p_i \neq p_{i-1}, \qquad 
\left| (t_i - t_{i-1}) - \frac{1}{2f} \right| < \delta,
\end{equation}
where \(\delta\) is a tolerance parameter accounting for temporal jitter. Events that do not satisfy this condition are discarded.

The PFF outputs a stream of valid modulation event pairs. 
For each accepted pair, we record its spatial coordinates \((x,y)\), its associated sub-aperture index \(r\) (region of interest, ROI), and its modulation cycle index \(k\). 
The set of all valid event pairs associated with ROI \(r\) in cycle \(k\) is denoted by \(\mathcal{P}_{r,k}\). These sets form the input to the centroid localization stage described below.

To obtain robust centroid estimates, we aggregate valid event pairs over a sliding window of consecutive modulation cycles. For each ROI \(r\) and output cycle \(k\), the aggregated event-pair set is defined as
\begin{equation}
\mathcal{Q}_{r,k} = \bigcup_{j=k-W+1}^{k} \mathcal{P}_{r,j},
\end{equation}
where \(W\) is the window length in modulation cycles.

The centroid of the focal-spot in ROI \(r\) at cycle \(k\) is then computed as the arithmetic mean of the event coordinates in \(\mathcal{Q}_{r,k}\):
\begin{equation}
\bar{x}_{r}[k] = \frac{1}{|\mathcal{Q}_{r,k}|} \sum_{i \in \mathcal{Q}_{r,k}} x_i,
\qquad
\bar{y}_{r}[k] = \frac{1}{|\mathcal{Q}_{r,k}|} \sum_{i \in \mathcal{Q}_{r,k}} y_i.
\end{equation}

For real-time operation, the cumulative sums required for centroid computation are updated incrementally as the sliding window advances, thereby avoiding redundant recomputation. A centroid estimate is reported only when the total number of valid event pairs in the window exceeds a predefined threshold \(T_{\text{min}}\); otherwise, the estimate for ROI \(r\) at cycle \(k\) is marked as invalid.
Additional implementation details of the algorithm are provided in Supplementary Section 4.

\begin{backmatter}
\bmsection{Funding} This research was supported by Zhejiang Provincial Natural Science Foundation of China under Grant No. LZ24F050003.


\bmsection{Conflict of Interest}
The authors declare no conflict of interest.

\bmsection{Permission to Reproduce Material}
No third-party material requiring permission is included in this manuscript.

\bmsection{Data Availability} Data underlying the results in Tables 1 and 2 are provided in Data 1 and Data 2, respectively.

\bmsection{Code Availability}
Upon publication, the source code and associated datasets will be made publicly available.

\bmsection{Supplemental Information} See Supplemental document for supporting content.

\end{backmatter}

\bibliography{sample}


\end{document}